\documentclass[aps,floatfix,twocolumn,showpacs]{revtex4}
\usepackage{graphics,epsfig}
\usepackage{graphicx}

\begin{document}

\title{Smooth transitions from Schwarzschild vacuum to de Sitter space}
\author{Steven Conboy and Kayll Lake \cite{email}}
\affiliation{Department of Physics, Queen's University, Kingston,
Ontario, Canada, K7L 3N6}
\date{\today}
\begin{abstract}
We provide an infinity of spacetimes which contain part of both
the Schwarzschild vacuum and de Sitter space. The transition,
which occurs below the Schwarzschild event horizon, involves only
boundary surfaces (no surface layers). An explicit example is
given in which the weak and strong energy conditions are satisfied
everywhere (except in the de Sitter section) and the dominant
energy condition is violated only in the vicinity of the boundary
to the Schwarzschild section. The singularity is avoided by way of
a change in topology in accord with a theorem due to Borde.
\end{abstract}
\maketitle

Without a quantum theory of gravity the endpoint of complete
gravitational collapse remains one of the central issues in
physics. The idea of regular black holes (those without internal
singularities) goes back at least forty years \cite{earlywork} and
the idea of a transition to de Sitter space across a  thin shell
in the Schwarzschild vacuum has seen extensive discussion in the
literature \cite{shells}. However, the introduction of a thin
shell is an approximation that allows properties of such a
transition to be absorbed into properties of the shell itself and
so it is certainly of interest to explore regular black holes
which do not contain thin shells. Relatively little work has been
done in this area \cite{boundary} and it is this type of
transition that is the subject of this work. Whereas our
considerations are classical (and formal), regular black holes are
of considerable interest from a non-classical viewpoint
\cite{nonclass}.

\bigskip

\begin{figure}[ht]
\epsfig{file=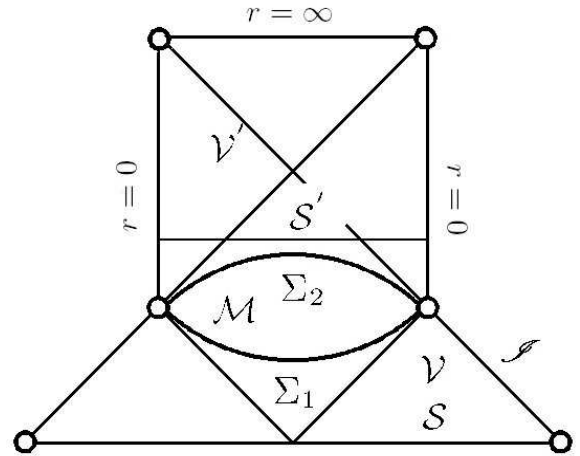,height=2.5in,width=3in,angle=0}
\caption{\label{image1}Global structure of part of the transition
from Schwarzschild ($\mathcal{V}$) to de Sitter space
($\mathcal{V}^{'}$) by way of boundary surfaces $\Sigma_{1}$ and
$\Sigma_{2}$. $\mathcal{M}$ is not vacuum. There is no singularity
because of the change in topology. The spacelike hypersurface
$\mathcal{S^{'}}$ is compact ($S^3$) but the spacelike
hypersurface $\mathcal{S}$ is not ($S^2 \times R$). The diagram
can be rotated about $\mathcal{S}$ to provide a geodesically
complete extension.}
\end{figure}

We consider the non-singular transition from the Schwarzschild
vacuum ($\mathcal{V}$) to de Sitter space ($\mathcal{V}^{'}$) by
way of spacelike boundary surfaces (not surface layers)
$\Sigma_{1}$ (below the horizon in Schwarzschild) and $\Sigma_{2}$
(above the cosmological horizon in de Sitter) through a non-vacuum
region $\mathcal{M}$. Part of the resultant maximally extended
spacetime is shown in FIG. \ref{image1}.

\bigskip

In this paper the singularity is avoided by way of a change in
topology in accord with a theorem due to Borde \cite{borde}. We
summarize the theorem here and refer the reader to Borde's papers
for details. Suppose that a future causally simple spacetime obeys
the null convergence condition (the Ricci tensor $R_{a}^{b}$ obeys
$R_{a}^{b}N^{a}N_{b} \geq 0$ for all null vectors $N^{a}$), is
null geodesically complete to the future and contains an
eventually future-trapped surface $\mathcal{T}$ (that is, the
divergence along each null geodesic that emanates orthogonally
from $\mathcal{T}$ becomes negative somewhere to the future of
$\mathcal{T}$). Then there is a compact slice to the causal future
of $\mathcal{T}$. As explained by Borde, it is the development of
this compact slice that allows singularity avoidance \cite{lake}.

\bigskip

In $\mathcal{M}$ write \cite{conventions} \cite{tsphere}
\begin{equation}
ds^2_{\mathcal{M}}=\frac{d\textrm{r}^2}{1-\frac{2M(\textrm{r})}{\textrm{r}}}+\textrm{r}^2
d\Omega^2+e^{2\Phi(\textrm{r})}dT^2,
\;\;\textrm{r}<2M(\textrm{r})\label{mmetric}
\end{equation}
where $d\Omega^2$ is the metric of a unit 2-sphere
($d\theta^2+sin(\theta)^2 d\phi^2$). Note that the metric
(\ref{mmetric}) is not static. (Whereas
$\xi^{\alpha}=\delta^{\alpha}_{T}$ is a hypersurface-orthogonal
Killing vector, it is spacelike.) For notational convenience, in
both $\mathcal{V}$ and $\mathcal{V^{'}}$ write
\begin{equation}
ds^2_{\mathcal{V}}=\frac{dr^2}{f(r)}+r^2 d\Omega^2-f(r)dt^2,
\;\;r<2m(r)\label{vmetric}
\end{equation}
where $f(r)\equiv {1-\frac{2m(r)}{r}}$. Again for notational
convenience, on $\Sigma_{1}$ and $\Sigma_{2}$ write
\begin{equation}
ds^2_{\Sigma}=R^2 d\Omega^2+d\lambda^2.\label{surface}
\end{equation}
Without loss in generality, we have taken $\theta$ and $\phi$
continuous but note that the coordinates in $\mathcal{V}$ and
$\mathcal{V^{'}}$ and on $\Sigma_{1}$ and $\Sigma_{2}$ are
otherwise distinct. The fact that the coordinates used in
(\ref{vmetric}) are valid only in a neighborhood of $\Sigma_{1}$
and $\Sigma_{2}$ in no way limits our analysis.

\bigskip

The continuity of the first and second fundamental forms at
$\Sigma_{1}$ and $\Sigma_{2}$ (the Darmois-Israel junction
conditions \cite{musgrave}) result in the conditions
\begin{equation}
R_{_{\Sigma}}=\textrm{r}_{_{\Sigma}}=r_{_{\Sigma}},\label{rsigma}
\end{equation}
\begin{equation}
M_{_{\Sigma}}=m_{_{\Sigma}},\label{msigma}
\end{equation}
and
\begin{equation}
\Phi^{'}=\frac{m^{*}r-m}{r^2(\frac{2m}{r}-1)}\label{Phisigma}
\end{equation}
at $\Sigma$, where $^{'}\equiv \frac{d}{d \textrm{r}}$ and
$^{*}\equiv \frac{d}{d r}$. These conditions must hold at both
$\Sigma_{1}$ and $\Sigma_{2}$. We note that since $m^{*}=0$ in
$\mathcal{V}$ then $\Phi^{'}<0$ at $\Sigma_{1}$. Further, since
$m=\frac{\Lambda r^3}{6}$ in $\mathcal{V}^{'}$, where $\Lambda$ is
the cosmological constant, $\Phi^{'}>0$ at $\Sigma_{2}$.

\bigskip

In the usual way, the energy density $\rho$ and principal
pressures $p_{1}$ and $p_{2}$ for $\mathcal{M}$ evaluate to
\begin{equation}
\tilde{\rho}=\frac{2}{\textrm{r}^3}(M+\Phi^{'}\textrm{r}(2M-\textrm{r})),\label{rho}
\end{equation}
\begin{equation}
\tilde{p_{1}}=-\frac{2M^{'}}{\textrm{r}^2},\label{p1}
\end{equation}
and
\begin{eqnarray}
\tilde{p_{2}}=-\frac{1}{\textrm{r}^3}((\Phi^{''}+(\Phi^{'})^2)\textrm{r}^2(2M-\textrm{r})\\\nonumber
+\Phi^{'}\textrm{r}(\textrm{r}M^{'}+M-\textrm{r})+\textrm{r}M^{'}-M),\label{p2}
\end{eqnarray}
where $\tilde{}\;$ indicates division by $8 \pi$.

\bigskip

From equations (\ref{rsigma}), (\ref{Phisigma}), (\ref{rho}) and
(\ref{p1}) it follows that
\begin{equation}
\tilde{\rho}_{_{\Sigma}}=\frac{2
m^{*}_{\Sigma}}{\textrm{r}_{\Sigma}^{2}}\label{rhosigma}
\end{equation}
and
\begin{equation}
\tilde{p_{1}}_{_{\Sigma}}=-\frac{2M^{'}_{\Sigma}}{\textrm{r}_{\Sigma}^{2}}.\label{p1sigma}
\end{equation}
From (\ref{rhosigma}) we then have
\begin{equation}
\tilde{\rho}_{_{\Sigma_{1}}}=0,\label{rhosigma1}
\end{equation}
and
\begin{equation}
\tilde{\rho}_{_{\Sigma_{2}}}=\Lambda.\label{rhosigma2}
\end{equation}
Equations (\ref{p1sigma}) and (\ref{rhosigma1}) make it clear that
the dominant energy condition \cite{poisson} is necessarily
violated in $\mathcal{M}$ at least in the neighborhood of
$\Sigma_{1}$ \cite{general}. Equations (\ref{p1sigma}) and
(\ref{rhosigma2}) make it clear that the dominant energy condition
need not be violated throughout $\mathcal{M}$ or at least in the
neighborhood of $\Sigma_{2}$. Equations (\ref{p1sigma}) and
(\ref{rhosigma1}) also show us that $M^{'}<0$ in $\mathcal{M}$ in
the neighborhood of $\Sigma_{1}$ for the weak (and null) energy
condition to be satisfied.

\bigskip

The degree to which energy conditions can be satisfied in
$\mathcal{M}$, subject to the boundary conditions specified,
depends on $\Phi(\textrm{r})$ and $M(\textrm{r})$. (In the de
Sitter section the strong energy condition is of course violated.)
Whereas one could demand isotropy of the pressure (and thereby
solve for $M(\textrm{r})$) we see no reason to do this here since,
as explained above, the dominant energy condition is necessarily
violated in at least a neighborhood of $\Sigma_{1}$ thus removing
at least part of $\mathcal{M}$ from normal classical physics.
Rather, we are interested in the degree to which the energy
conditions must be violated and we content ourselves here with a
simple example so as to explore this.

\bigskip

\begin{figure}[ht]
\epsfig{file=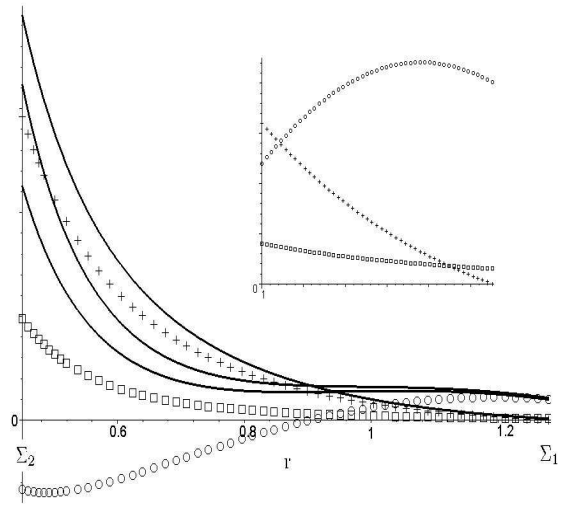,height=2.7in,width=3in,angle=0}
\caption{\label{image2} The energy conditions in $\mathcal{M}$.
The energy density $\tilde{\rho}$ is indicated by a cross,
$\tilde{p_{1}}$ by a box and $\tilde{p_{2}}$ (which becomes a
tension) by a circle. The solid curves give (top down at
$\textrm{r}=0.6$) $\tilde{\rho}+\tilde{p_{1}}$,
$\tilde{\rho}+\tilde{p_{1}}+\tilde{p_{2}}$ and
$\tilde{\rho}+\tilde{p_{2}}$. Detail near $\Sigma_{1}$ where the
dominant energy condition fails is shown in the insert. }
\end{figure}

Take
\begin{equation}
\Phi(\textrm{r})=(1-\textrm{r}^2)(\textrm{r}^2-2)\label{phispecial}
\end{equation}
and
\begin{equation}
M(\textrm{r})=\frac{2}{\textrm{r}}.\label{mspecial}
\end{equation}
It follows that $\textrm{r}_{_{\Sigma_{1}}}$ is uniquely
determined ($ \sim 1.276$) as is $\textrm{r}_{_{\Sigma_{2}}}$ ($
\sim 0.451$) and therefore $\Lambda$
($=12/\textrm{r}_{\Sigma_{2}}^{4}$). The energy conditions are
shown in FIG.  \ref{image2} where it can be seen that the weak,
null and strong energy conditions hold for
$\textrm{r}_{_{\Sigma_{2}}} < \textrm{r} <
\textrm{r}_{_{\Sigma_{1}}}$ and even the dominant energy condition
holds for $\textrm{r}_{_{\Sigma_{2}}} < \textrm{r} < \sim  0.8 \;
\textrm{r}_{_{\Sigma_{1}}}$.

Clearly there is an infinity of spacetimes which contain part of
both the Schwarzschild vacuum and de Sitter space with the
transition (below the Schwarzschild event horizon) involving only
boundary surfaces (no surface layers). The simple example given
above shows that it is not difficult to find examples in which the
weak, null and strong energy conditions hold throughout (excepting
of course in the de Sitter section) and the violation of the
dominant energy condition is isolated. The singularity is avoided
by way of a change in topology in accord with a theorem due to
Borde \cite{borde}.

\begin{acknowledgments}
This work represents part of an undergraduate thesis (of SC) and
was supported in part by a grant from the Natural Sciences and
Engineering Research Council of Canada (to KL). We thank Jos\'{e}
Senovilla for comments which helped to clarify our presentation.
Portions of this work were made possible by use of
\textit{GRTensorII} \cite{grt}.
\end{acknowledgments}


\begin{thebibliography}{}\label{sec:TeXbooks}
\bibitem[*]{email}{Electronic Address: lake@astro.queensu.ca}
\bibitem{earlywork} See A. D. Sakharov, Sov. Phys. JETP {\bf 22}, 241 (1966)
and E. B. Gliner, Sov. Phys. JETP {\bf 22}, 378 (1966) who both
considered de Sitter space as the final state of gravitational
collapse. J. M. Bardeen (GR-5 Book of Abstracts, Tbilisi (1968))
prsented a modification of the Reissner-Nordstr\"{o}m solution
with an event horizon but no singularities. It is important to
note that Bardeen was well aware of the importance of topology
change in his example. This has been discussed in detail by A.
Borde, Phys. Rev. D \textbf{50},
 3692 (1994)  {\tt arXiv:gr-qc/9403049}
\bibitem{shells} See M. A. Markov, JETP Lett. {\bf 36}, 265 (1982) and Ann. Phys. {\bf
155}, 333 (1984), M. R. Bernstein, Bull. Amer. Phys. Soc.
{\bf16},1016 (1984), V. P. Frolov, M. A. Markov and V. F.
Mukhanov, Phys. Lett. B \textbf{216}, 272 (1989), K. Lake and T.
Zannias, Phys. Lett. A \textbf{140}, 291 (1989), V. P. Frolov, M.
A. Markov and V. F. Mukhanov, Phys. Rev. D \textbf{41}, 383
(1990), R. Balbinot and E. Poisson, Phys. Rev. D \textbf{41}, 395
(1990), D. Morgan, Phys. Rev. D \textbf{43}, 3144 (1991), M.
Visser and D. Wiltshire, Class. Quant. Grav. \textbf{21}, 1135
(2004) {\tt arXiv:gr-qc/0310107} For a related discussion see E.
Poisson and W. Israel, Class. Quant. Grav. {\bf 5} (1988) L201.
For related discussions of lightlike boundaries see W. Shen and S.
Zhu, Phys. Lett. A \textbf{126}, 229 (1988), C. Barrab\`es and V.
P. Frolov, Phys. Rev. D \textbf{53}, 3215 (1996) {\tt
arXiv:hep-th/9511136} and Helv. Phys. Acta \textbf{69}, 253 (1996)
{\tt arXiv:gr-qc/9607040}
\bibitem{boundary} M. Mars, M. M. Mart\'\i n-Prats and J. M. M. Senovilla,
Class.\ Quant. Grav. \textbf{13}, L51 (1996) have considered
regular Schwarzschild black holes (without transition to de Sitter
space). See J. M. M. Senovilla, Gen. Rel. Grav. \textbf{30}, 701
(1998) for further discussion. I. Dymnikova, Gen. Rel. Grav.
\textbf{24}, 235 (1992) (see also I. Dymnikova and E. Galaktionov
{\tt arXiv:gr-qc/0409049}) has considered a form of
(\ref{vmetric}) which is de Sitter for $r \rightarrow 0$ and
Schwarzschild for $r \rightarrow \infty$. See also E. Elizalde and
S. R. Hildebrandt, Phys. Rev. D \textbf{65} 124024 (2002) {\tt
arXiv:gr-qc/0202102} who consider regular interiors with a
prescribed equation of state.
\bibitem{nonclass}The limiting curvature hypothesis (see
\cite{shells}) has been given a non-classical implementation (V.
Mukhanov and R. Brandenberger, Phys. Rev. Lett. {\bf 68}, 1969
(1992) and R. Brandenberger, V. Mukhanov and A. Sornborger, Phys.
Rev. {\bf D48}, 1629 (1993)) leading to non-classical regular
black hole constructions in 1+1 dimensions (M. Trodden, V.
Mukhanov and R. Brandenberger, Phys. Lett.  {\bf B316}, 483
(1993)). There are many current discussions of non-classical
regular black holes in various contexts. See, for example, L.
Modesto,  Phys. Rev. D \textbf{70} 124009 (2004) {\tt
arXiv:gr-qc/0407097} V. Husain and O. Winkler ``Quantum resolution
of black hole singularities" {\tt arXiv:gr-qc/0410125} T. Hirayama
and B. Holdom, Phys. Rev. D \textbf{68} 044003 (2003) {\tt
arXiv:hep-th/0303174} D. A. Easson, J. High Energy Phys.
JHEP02(2003)037 {\tt arXiv:hep-th/0210016} D. A. Easson, R. H.
Brandenberger, J. High Energy Phys. JHEP06(2001)024, {\tt
arXiv:hep-th/0103019}
\bibitem{borde}A. Borde, Phys. Rev. D \textbf{55}, 7615 (1997) {\tt
arXiv:gr-qc/9612057 } See also A. Borde, Phys. Rev. D \textbf{50},
3692 (1994) {\tt arXiv:gr-qc/9403049 }
\bibitem{lake} If the initial spacetime is not asymptotically flat
then more elaborate constructions are possible. See K. Lake (in
preparation).
\bibitem{conventions}We use geometrical units and a signature of
$+2$. For convenience, explicit functional dependence is usually
shown only on the first appearance of a function.
\bibitem{tsphere}This construction, though unusual, is by no means
new. In the older Russian literature this is known as a
``T-sphere". The work on these configurations by I. D. Novikov and
V. A. Ruban has recently been reprinted. See the Editor's notes in
Gen. Rel. Grav. \textbf{33}, 369 and 2255 (2001) and the articles
accompanying these notes. Some further discussion (due to K. S.
Thorne) is given in section 3.2 of Ya. B. Zel'dovich and I. D.
Novikov, \textit{Relativistic Astrophysics Volume 1 Stars and
Relativity }(The University of Chicago Press, Chicago, 1971). See
also G. C. McVittie and R. J. Wiltshire, Int. Jour. Theor. Phys.
\textbf{14}, 145 (1975). The concept of ``T" and ``R" regions can
be distinguished invariantly as follows: Define the vector field
$k$ by
\begin{eqnarray}
k_{\;a} \equiv \nabla_{a} (C_{b c d e} C^{b c d e})=-\nabla_{a}
(\bar{C}_{b c d e} \bar{C}^{b c d e}) \nonumber
\end{eqnarray}
where $C_{b c d e}$ is the Weyl tensor and $\bar{C}_{b c d e}$ its
dual tensor. In what follows we assume that  $C_{b c d e}{C}^{b c
d e} \neq 0$ except perhaps locally. A  ``T" region of spacetime,
defined by $k$, corresponds to timelike $k$ and an ``R" region of
spacetime corresponds to spacelike $k$. These are invariant
algorithmic generalizations of the classical notions of ``R" and
``T" regions and reduce to them in the spherically symmetric case.
\bibitem{musgrave} See, for example, P. Musgrave and K. Lake, Class. Quant. Grav.
\textbf{13}, 1885 (1996).
\bibitem{poisson} For a discussion of energy conditions see S. W. Hawking and G. F. R. Ellis, \textit{The Large Scale Structure of Space-Time}
(Cambridge University Press, Cambridge, 1973), M. Visser,
\textit{Lorentzian Wormholes} (Springer-Verlag, New York, 1996),
E. Poisson, \textit{A Relativist's Toolkit: The Mathematics of
Black-hole Mechanics} (Cambridge University Press, Cambridge,
2004).
\bibitem{general} The necessary violation of the dominant energy condition is a very general
result. See, for example, Mars et al. \cite{boundary}.
\bibitem{grt}This is a package which runs within Maple. It is entirely
distinct from packages distributed with Maple and must be obtained
independently. The GRTensorII software and documentation is
distributed freely on the World-Wide-Web from the address \textit{
http://grtensor.org}
\end{thebibliography}
\end{document}